\documentstyle[12pt,aaspp4]{article}

\newcommand{\etal}{{\it et al.\/}}

\newcommand{\msun}{M$_{\odot}$}
\def\vi{\hbox{$V\!-\!I$}} 

\newcommand{\hst}{{\it HST\/}} 
\newcommand{\sn}{$S_N$\/}
\newcommand{\kms}{km~sec$^{-1}$}
\newcommand\cola {\null}

\newcommand\colc {&}
\newcommand\cold {&}
\newcommand\cole {&}

\newcommand\colg {&}

\newcommand\coli {&}
\newcommand\eol{\\}
\newcommand{\ctlea}{\multicolumn{1}{c}}

\begin{document}

\title{The Specific Globular Cluster Frequencies of Dwarf Elliptical
Galaxies from the {\it Hubble Space Telescope}\footnote{Based on 
observations with the NASA/ESA
{\it Hubble Space Telescope}, obtained at the Space Telescope Science
Institute, which is operated by the Association of Universities for
Research in Astronomy, Inc., under NASA contract No.  NAS5-26555}}

\author{ Bryan W. Miller\footnote{Sterrewacht Leiden, Postbus 9513, 2300 
RA Leiden, The Netherlands}, Jennifer M. Lotz\footnote{Dept. of Physics 
and Astronomy, Johns Hopkins University, Baltimore, 
MD, 21218}, Henry C. Ferguson\footnote{Space Telescope 
Science Institute, 3700 San Martin Dr., Baltimore, MD 21218}, 
 Massimo Stiavelli$^4$, \\ and Bradley C. Whitmore$^4$}

\centerline{To appear in the Astrophysical Journal Letters}

\begin{abstract}
The specific globular cluster frequencies (\sn) for 24 dwarf
elliptical (dE) galaxies in the Virgo and Fornax Clusters and the Leo
Group imaged with the {\it Hubble Space Telescope} are presented.
Combining all available data, we find that for nucleated dEs --- which
are spatially distributed like giant ellipticals in galaxy clusters ---
$\bar{S}_N({\rm dE,N})=6.5\pm1.2$ and \sn\ increases with $M_V$, while
for non-nucleated dEs --- which are distributed like late-type
galaxies --- $\bar{S}_N({\rm
dE,noN})=3.1\pm0.5$ and there is little or no trend with $M_V$.  The
$S_N$ values for dE galaxies are thus on average significantly higher
than those for late-type galaxies, which have $S_N \lesssim 1$.  This
suggests that dE galaxies are more akin to giant Es than to late-type
galaxies. If there are dormant or stripped irregulars hiding among the
dE population, they are likely to be among the non-nucleated
dEs. Furthermore, the similarities in the properties of the globular
clusters and in the spatial distributions of dE,Ns and giant Es
suggest that neither galaxy mass or galaxy metallicity is responsible for
high values of \sn.  Instead, most metal-poor GCs may have formed in
dwarf-sized fragments that merged into larger galaxies.
\end{abstract}


\section{Introduction}

There are several distinct classes of galaxies with smooth surface
brightness profiles and absolute magnitudes fainter than $M_B = -18$.
Brighter than $M_B = -15$, most dwarf elliptical (dE) galaxies have
central nuclei which are unresolved from the ground at the distance
of the Virgo Cluster. However, there are quite a few dEs in this
luminosity range that have no sign of a nucleus.  Within the Virgo and
Fornax clusters, the non-nucleated dEs form an extended population,
with a spatial distribution similar to the spirals and irregulars
rather than the bright ellipticals and the rest of the dEs (Ferguson
\& Sandage \markcite{FS89} 1989).  A possible explanation for the
difference is that the non-nucleated dEs are stripped or harassed
dwarf irregular (dI) galaxies (e.g. Lin \& Faber \markcite{LF83} 1983;
Moore \etal \markcite{Moore 98} 1998).  The most popular alternative
scenario for dE formation (e.g.  Dekel \& Silk \markcite{DS86} 1986) is
that they formed in one monolithic collapse, and subsequently ejected
their ISM via supernova winds.  To distinguish these scenarios we
require an observational probe of the earliest and most important
episodes of star formation in dwarf galaxies.  Globular cluster (GC)
systems may provide such a probe.

The specific globular cluster frequency (\sn, the number
of clusters normalized to a galaxy with $M_V=-15$) is correlated with
Hubble type such that it increases from late type spirals to early
type spirals and ellipticals, suggesting a correlation with the size
of the bulge component (Harris \markcite{h91araa} 1991a).  The empirical
variation of \sn\ from early to late types suggests a simple test for
whether dEs are more closely related to giant ellipticals (Es) or to
irregulars (Harris \markcite{h91pasp} 1991b). If part of the E family,
they should have $S_N \sim 2-5$; if part of the late-type disk galaxy
family, we expect $S_N \lesssim 1$.  Finally, one might expect that
nucleated dEs will have higher values of $S_N$ than non-nucleated dEs.

We have begun a {\it Hubble Space Telescope} snapshot survey in order
to study the properties of the GCs and nuclei of dE galaxies.  In this
paper we present results on \sn\ for 24 galaxies in the Virgo and
Fornax Clusters and the Leo Group.  Table~1 gives the galaxy names
morphological types from photographic catalogs.  This sample
represents all the data suitable for determining $S_N$ taken during
\hst\ Cycle~6.  We assume distance moduli of 31.2 for the Virgo
Cluster, 31.4 for the Fornax Cluster, and 30.3 for the Leo Group.  When 
presenting an uncertainty of an average quantity we quote the 
standard deviation of the mean. The
observations and analysis are summarized in Section~2.  We present the
results on the specific frequencies in Section~3 and discuss some
implications in Section~4.  The data will be more fully presented
in Miller \etal \markcite{de6data} (1998, Paper~II).

\section{Observations and Data Reduction}

WFPC2 snapshot images are taken in the F555W ($2 \times 230$ sec) and
F814W (300 sec) bandpasses with the galaxies centered in chip WF3.
The procedures for object detection and classification, aperture
photometry, and photometric calibration are similar to those used by
Miller \etal \markcite{n7252} (1997).  From these data we can determine
the (\vi) colors of GC candidates down to a completeness limit of
$V\approx25$.

The specific globular cluster frequency, \sn, is defined as $S_N =
N_{c}10^{0.4(M_V +15)}$ where $N_c$ is the total number of clusters
(Harris \& van den Bergh \markcite{HvdB81} 1981). Cluster candidates are
selected based on color ($0.5 < V\!-\!I < 1.5$), size (FWHM $<$ 2.5
pixels), and accuracy of photometry (err(\vi) $<$ 0.3~mag). The net
number of cluster candidates for a galaxy, $N_c$, is the total number
of cluster candidates in WF3 minus the expected number of background
and foreground objects.  The number density of background objects are
determined from the objects in WF2 and WF4.  We must also correct for
undetected faint clusters.  The luminosity function of all the cluster
candidates is consistent with a Gaussian with $M_V^{peak}\approx -7.3$
and $\sigma\approx1.0$. Tests then show that we are detecting 85--90\%
of the clusters in Virgo and Fornax (down to $M_V\approx-6$) and
$\sim98\%$ in Leo (down to $M_V\approx-5$).  Therefore, we multiply
the net number of detected GCs for Virgo and Fornax galaxies by 1.15
to correct for this incompleteness.

We also measure the corresponding galaxy luminosity in WF3. After sky
subtraction, most point sources and resolved background galaxies are
removed from the image.  The only cluster candidates which are not
excised are the nuclei with $M_V<-10.5$ (see below).  Then, the total
counts remaining in the image are summed and corrected for foreground
extinction to determine $M_V^0$.
Uncertainties range between 0.1~mag for the brightest systems to
$\sim0.4$~mag for the faintest.

An important problem is the classification of galaxies into nucleated
and non-nucleated samples.  Our first criterion for
classifying nuclei is the projected radial distance, $R_{\rm proj}$,
from the galaxy center as defined by the centroid of the light
distribution after removing any central point-like sources. The
uncertainties in the central coordinates are determined by measuring
centroids with a range of centering-box widths.  These uncertainties
yield the number of standard deviations, $\sigma_{\rm proj}$, that the central
objects are offset from the centers.  As a further test, we have used
Monte-Carlo simulations to compute the probability, $P$, that an
object as bright as the central cluster candidate would be found
within $R_{\rm proj}$ of the center by chance.  In each trial for a
given galaxy, $N_c$ artificial clusters are drawn from a Gaussian
luminosity function and are given
the same exponential radial distribution as measured for that
galaxy.  Many trials are run to reduce uncertainties due to small
numbers.  We classify a galaxy as nucleated if the center-most cluster
candidate is within 3$\sigma_{\rm proj}$ of the galaxy center and
$P\lesssim 1$\%.

In most cases the nucleus turns out to be the most luminous GC, but in
a few galaxies it is significantly brighter than the typical GC.  With
M$_V\lesssim-10.5$ these ``clusters'' are brighter than any globular
cluster in the Galaxy or M31, and comparable to the brightest in M87
(Djorgovski \markcite{Djorgovski 93} 1993; Battistini \etal
\markcite{BBBFFMB87} 1987; Whitmore \etal \markcite{WSLMB95} 1995). These
objects may be more like true nuclei.  It is unclear whether nuclei of
large galaxies are kinematically and structurally distinct from star
clusters or simply giant versions of the latter (Kormendy \& McClure
\markcite{KM93} 1993; Carollo \etal \markcite{Carollo 97} 1997).  To
be conservative we have chosen not to count nuclei with M$_V<-10.5$ as
clusters for the purposes of calculating \sn.

\section{Results}

\begin{figure}
\plotfiddle{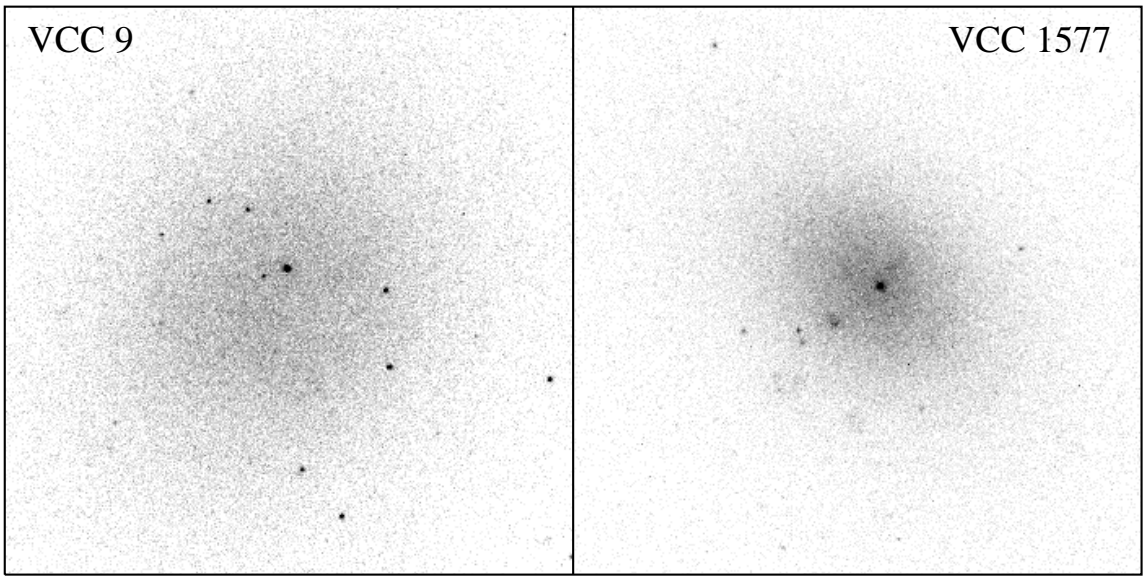}{5in}{90}{70}{70}{275}{0}
\caption{The central 35\arcsec\ of two dE galaxies with possible
off-center nuclei.  Each panel shows a physical region about 3~kpc
across. \label{fig:offsetn}}
\end{figure}

Table~1 gives our measured values of $M_V^0$, $N_c$ (corrected for the
faint end of the LF), $S_N$, $R_{\rm proj}$, $\sigma_{\rm proj}$ by
which $R_{\rm proj}$ differs from the galaxy center, and probability
$P$ of finding the center-most cluster candidate within $R_{\rm proj}$
by chance for 24 dE galaxies.  The galaxies have been split into
nucleated and non-nucleated categories based on the criteria mentioned
above.  In two cases our classification differs from the ground-based
classification.  FCC~324 was originally in our non-nucleated sample
but we find that it is nucleated.  VCC~9 has a catalog type of dE1N.
Its most luminous cluster is close to the center, but with
$R_{\rm proj}=155$~pc it is too offset to meet our criteria for being
nucleated.   Two other
galaxies may be intermediate types.  VCC~1577 and FCC~48 have their
most luminous GC within 3$\sigma$ of their centers, but these galaxies do
not fulfill the probability test.  The central regions of VCC~9 and VCC~1577 
are shown in Figure~\ref{fig:offsetn}.

We have one galaxy, VCC~1254, in common with the
ground-based study of Durrell \etal \markcite{DHGP96} (1996).  
Adjusting their value to our distance
modulus yields $S_N=6.4\pm3.0$.  Our result from Table~1 is 
$S_N=6.2\pm2.2$ and agrees very well with their measurement.  Therefore,
our method appears reliable.

Figure~\ref{fig:snmv} shows log(\sn) versus $M_V$ for all dE galaxies
with measured \sn.  In the figure, galaxies with $S_N=0$ have been
assigned $\log(S_N)=0.0$.  There is a clear trend of increasing $S_N$
with decreasing galaxy luminosity for the nucleated galaxies (filled
symbols).  A weighted least-squares fit to the nucleated galaxies with
$M_V<-13.5$ gives a slope of $0.24\pm0.04$ (solid line).  The
non-nucleated galaxies (open and half-filled symbols) do not show such
a trend, a fit these points gives a
slope of $0.10\pm0.05$ (dashed line).  The trend for the nucleated
galaxies was seen by Durrell \etal \markcite{DHGP96} (1996) but our larger
sample now indicates that there is a difference in \sn\ between
nucleated and non-nucleated dEs.

At magnitudes fainter than $M_V=-14$ the large fluctuations in \sn\
caused by small number statistics may blur the distinction between
nucleated and non-nucleated galaxies.  An extension of the fit to the
nucleated galaxies in Figure~\ref{fig:snmv} passes through the points
for both the Fornax and Sagittarius dwarfs, the dEs with the highest
measured \sn.  Even though the total extent of Sagittarius is
uncertain because of tidal disruption by the Galaxy, M~54 is generally
thought to be its nucleus (Sarajedini \& Layden \markcite{SL95} 1995).
Fornax is usually considered non-nucleated, but it has a high \sn\
like the other dE,Ns and it has a metal-rich GC only $\sim\!150$~pc from
its center.  The triangle labeled ``LG dSph'' represents the $\sim12$
non-nucleated Local Group dwarf spheroidals with $S_N=0$. For now, it
may be best to think of the dE,Ns as defining the upper envelope to
the trend of increasing \sn\ with $M_V$.

\section{Discussion}

\begin{figure}
\plotone{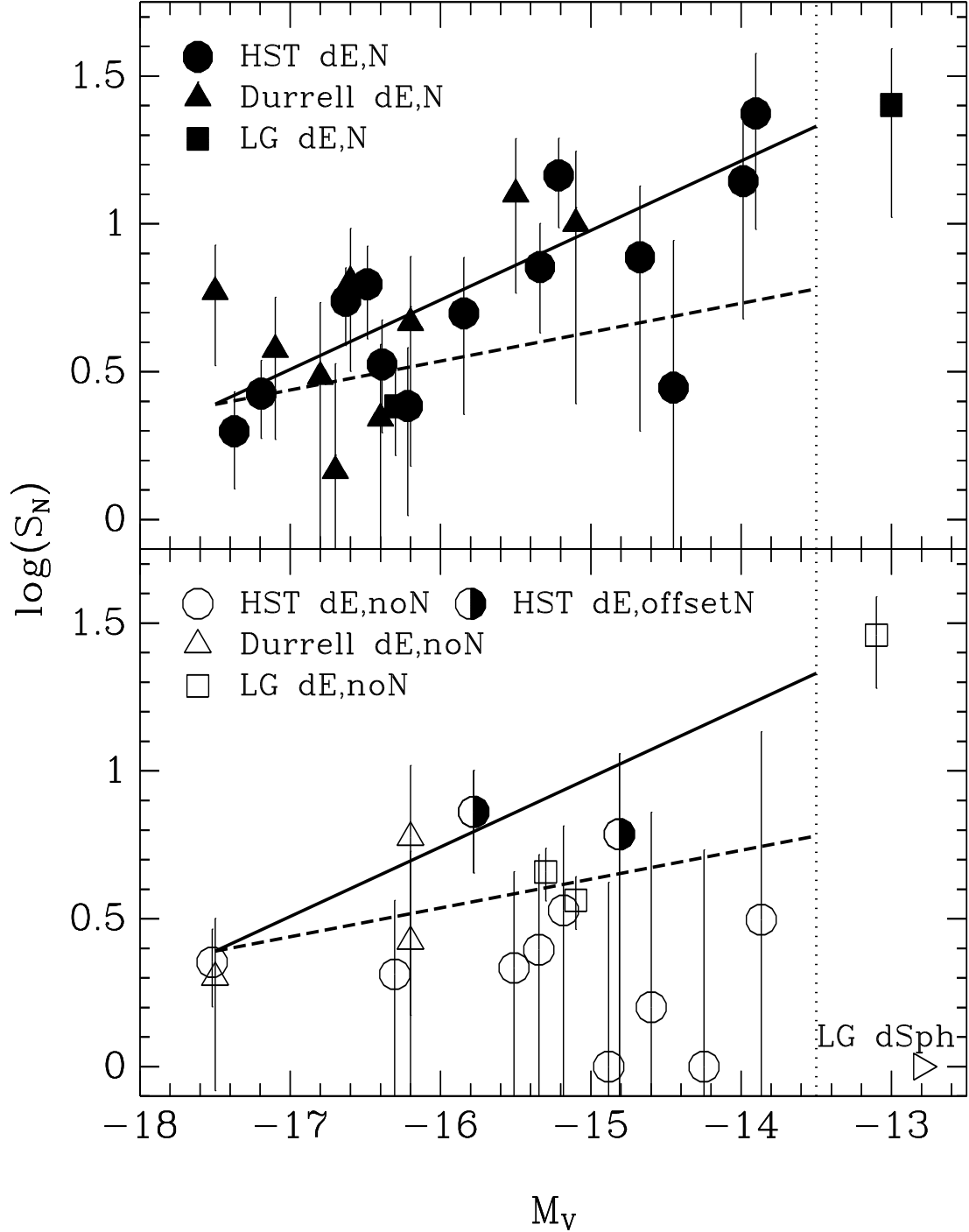}
\caption{Trends of log(\sn) with $M_V$ for all dE galaxies with
measured $S_N$.  Circles are from the current work, triangles are from
Durrell \etal\ (1996) after adjusting to our assumed distances, and
squares are from the Local Group.  The solid line gives the weighted
fit to the nucleated galaxies (filled symbols) and has a slope $0.24
\pm 0.04$.  The dashed line gives the fit to the galaxies with offset
nuclei (FCC~48 and VCC~1577) and no nuclei (open symbols) and has a
shallower slope of $0.10 \pm 0.05$. The dotted vertical line shows
the magnitude cutoff for the fits at $M_V=-13.5$.\label{fig:snmv}}
\end{figure}

High resolution \hst\ images of a large sample of dE galaxies provide
an improved estimate of their specific globular cluster frequency, and
an indication that \sn\ depends on galaxy type and luminosity. We
confirm earlier ground-based estimates of a high $S_N$ for dE galaxies
(Durrell \etal\ \markcite{DHGP96} 1996).  Adjusted to our distances
their sample of 9 dE,N and 3 dE,noN galaxies gives a unweighted mean
of $\bar{S}_N = 5.1\pm1.0$. For comparison, our sample gives
$5.3\pm1.1$. These values are consistent, and are significantly higher
than the mean for late type galaxies $\bar{S}_N = 0.5\pm0.2$ (Harris
\markcite{h91araa} 1991a).  In addition, we find for our sample that
$\bar{S}_N({\rm dE,N})=7.5\pm1.8$ and $\bar{S}_N({\rm
dE,noN})=2.8\pm0.7$. The difference in $S_N$ between dE,N and dE,noN
in Durrell \etal's sample is much less apparent, probably due to
smaller sample size.  The combined sample with $M_V<-13.5$ shown in
Figure~\ref{fig:snmv} yields $\bar{S}_N({\rm dE,N})=6.5\pm1.2$ and
$\bar{S}_N({\rm dE,noN})=3.1\pm0.5$; the \sn\ for nucleated dEs is a
factor of two higher than for non-nucleated dEs.

This result is a further argument against the formation of dE
galaxies {\it as a class} via the quiescent evolution
of dI galaxies.  The statistics of $S_N$ in dIs
 are not very good, but typical values
appear to around $S_N = 0.5$ (Harris \markcite{h91araa}
1991a).  Simple removal of gas from a dI galaxy
would result in modest fading. For example, consider a galaxy that forms
stars at a constant rate for 5~Gyr, whereupon it is stripped
of gas and ceases forming stars. In the 5~Gyr after star
formation ends the galaxy fades by $\sim\!1.5$ mag in $V$. So, if it
started with $S_N = 0.5$, it would end up with $S_N \approx 2.0$,
which is at the low end of the observed range for the dEs.  
A longer period of fading would result in higher $S_N$, but 
the resulting surface brightness would tend to be fainter than
measured in bright dE,Ns. Similarly, a model in which dEs {\it in
general} represent the quiescent periods between stochastic bursts of
star formation in dIs (Tyson \& Scalo \markcite{TS88} 1988) appears
inconsistent with the high average value of $S_N$ measured for dE
galaxies.  The conclusion is that {\it most} bright dE galaxies, and
the dE,Ns in particular, are not simply faded or dormant dI galaxies.

The low mean $S_N$ among {\it non-nucleated} dE galaxies suggests that
any faded dIs that exist in the cluster populations are likely part of
this class.  A number of the dE,noN galaxies in our sample have
$S_N<1$, consistent with the values measured in dIs.  Also, our
calculation above shows that {\it faded} dIs would have $S_N$ close to
the mean $S_N$ that we measure for the dE,noN class. A further
argument in favor of this possibility is the observation that the
non-nucleated dEs brighter than $M_B = -14$ have an extended spatial
distribution in the Virgo and Fornax Clusters like the spatial
distribution of spirals and irregulars. It would be interesting to
search for evidence of gas either in or around some of the brighter
non-nucleated dE galaxies in these clusters.  Also, stripped dIs may
still exhibit rotation.

The colors and luminosity functions of the GCs in dEs are
also similar to the old, metal-poor GC populations in giant
ellipticals (Paper~II).  Therefore, GCs in both giant Es and dEs seem
to have formed at similar times, with the same metallicities, and with
the same efficiencies.  It is not obvious why this
should be so. The velocity dispersions of typical giant
ellipticals are more than 250~\kms, while for  dE
galaxies typical velocity dispersions are likely to be less
than 100~\kms\ (Bender \& Nieto \markcite{BN90} 1990). Thus it
appears that galaxy velocity dispersion (and thus potential well
depth) is not a key parameter.
In addition, the metallicities of our dE galaxies, if adequately
represented by their colors, are probably less than half solar, while
the metallicities for many giant ellipticals exceed solar. Thus, the
current metallicity of the galaxy also does not seen to play an
important role in driving $S_N$.  How can such similar GCs form in
such different parent galaxies?

This can be explained if most GCs are formed in dwarf galaxy sized
``fragments'' that latter merge into giant ellipticals (Searle \& Zinn
\markcite{SZ78} 1978; Harris \& Pudritz \markcite{HP94} 1994). The
fact that E galaxies and dE,N galaxies inhabit the densest regions of
clusters suggests that the cluster environment is somehow
important for GC formation, for example through stronger interactions,
higher external pressures, or higher star formation rates during the
epoch of cluster formation (McLaughlin \&
Pudritz \markcite{MP96} 1996; Elmegreen \& Efremov \markcite{EE97} 1997;
Harris, Harris, \& McLaughlin \markcite{HHM98} 1998).  These
environments would also be prone to the most merging.  Additional
evidence for this type of scenario is the accretion of the GCs from
the Sagittarius dwarf into the halo of the Galaxy. Thus, accretion of
GC-rich dwarfs may be a natural explanation of the metal-poor GCs in giant
ellipticals and spirals.
        
Turning to the trend with luminosity, our data suggests that, at least
among nucleated dEs, less luminous galaxies have higher $S_N$. We have
investigated several possible explanations for this phenomenon.
First, winds may be more efficient at quenching star formation in the
lower luminosity dEs.  Second, many of the original clusters may have
merged into the bright nucleus.  A $10^6$~\msun\ cluster moving at
50~\kms\ through an isothermal halo will decay from a radius of 3~kpc
in less than a Hubble time (Binney \& Tremaine \markcite{BT87} 1987).
We have calculated that \sn\ could decrease by a factor of about 1.5
if the nuclei brighter than $M_V=-10.5$ are merged GCs.  However,
these nuclei do not appear to be simple conglomerates of GCs.  They
are on average $0.06\pm0.02$~mag redder in $(V-I)$ than the mean color
of the GCs in those galaxies.  If this is due to metallicity (Couture
\etal \markcite{CHA90} 1990), then the nuclei are about 0.3~dex more
metal rich than the GCs. Thus, additional nuclear star formation,
perhaps as a result of interactions, may have
occurred (also see Durrell \etal \markcite{DHGP96} 1996).
       
In summary, we have confirmed earlier estimates of a high
globular cluster specific frequency in dE galaxies in nearby
clusters of galaxies. Further, we have shown evidence for a difference in
$S_N$ between nucleated and non-nucleated dEs, and for a 
trend in $S_N$ with luminosity (at least for the dE,N galaxies).
The results reinforce the status of dE,N galaxies as the likely
low-luminosity tail of the giant-E galaxy sequence, and suggest
that the processes that formed both types of galaxies were
related and probably influenced by the cluster environment. 
The present trends with luminosity and type are tentative and
must be confirmed with larger samples. Fortunately, these 
are relatively inexpensive observations to make with \hst, and
we are engaged in a extension of the survey that will observe
up to 30 more faint dEs and up to 30 dIs to improve the
statistical comparisons.
        
\acknowledgements

We would like to thank Bill Harris, Fran\c{c}ois Schweizer,  
and Tim de~Zeeuw for their comments and Jen Mack for help
with the surface photometry.  This research has made use of NASA's
Astrophysics Data System Abstract Service.  Support for this work was
provided by NASA through grant number GO-6352 from the Space Telescope
Science Institute, which is operated by AURA, Inc., under NASA
contract NAS5-26555.

\clearpage

\begin{thebibliography}{}

\bibitem[\protect\citeauthoryear{{Battistini} et~al.}{{Battistini}
  et~al.}{1987}]{BBBFFMB87}
{Battistini}, P.~L., {Bonoli}, F., {Braccesi}, A., {Federici}, L.,
  {Fusi-Pecci}, F., {Marano}, B.,  \& {Borngen}, F. 1987, A\&AS, 67, 447

\bibitem[\protect\citeauthoryear{{Bender} \& {Nieto}}{{Bender} \&
  {Nieto}}{1990}]{BN90}
{Bender}, R.,  \& {Nieto}, J.-L. 1990, A\&A, 239, 97

\bibitem[\protect\citeauthoryear{{Binney} \& {Tremaine}}{{Binney} \&
  {Tremaine}}{1987}]{BT87}
{Binney}, J.,  \& {Tremaine}, S. 1987, Galactic Dynamics (Princeton: Princeton
  Univ. Press)

\bibitem[\protect\citeauthoryear{{Carollo} et~al.}{{Carollo}
  et~al.}{1997}]{CSdM97}
{Carollo}, C.~M., {Stiavelli}, M., {de~Zeeuw}, P.~T.,  \& {Mack}, J. 1997, AJ,
  114, 2366

\bibitem[\protect\citeauthoryear{{Couture}, {Harris}, \& {Allwright}}{{Couture}
  et~al.}{1990}]{CHA90}
{Couture}, J., {Harris}, W.~E.,  \& {Allwright}, J. W.~B. 1990, ApJS, 73, 671

\bibitem[\protect\citeauthoryear{{Dekel} \& {Silk}}{{Dekel} \&
  {Silk}}{1986}]{DS86}
{Dekel}, A.,  \& {Silk}, J. 1986, ApJ, 303, 39

\bibitem[\protect\citeauthoryear{{Djorgovski}}{{Djorgovski}}{1993}]{Djorgovski%
93p373}
{Djorgovski}, S.~G. 1993, in Structure and Dynamics of Globular Clusters, ed.
  S.~G. {Djorgovski} \& G.~{Meylan} (San Francisco: ASP), 373

\bibitem[\protect\citeauthoryear{{Durrell} et~al.}{{Durrell}
  et~al.}{1996}]{DHGP96}
{Durrell}, P., {Harris}, W.~E., {Geisler}, D.,  \& {Pudritz}, R. 1996, AJ, 112,
  972

\bibitem[\protect\citeauthoryear{{Elmegreen} \& {Efremov}}{{Elmegreen} \&
  {Efremov}}{1997}]{EE97}
{Elmegreen}, B.~G.,  \& {Efremov}, Y.~N. 1997, ApJ, 480, 235

\bibitem[\protect\citeauthoryear{{Ferguson} \& {Sandage}}{{Ferguson} \&
  {Sandage}}{1989}]{FS89}
{Ferguson}, H.~C.,  \& {Sandage}, A. 1989, ApJ, 346, L53

\bibitem[\protect\citeauthoryear{{Harris}}{{Harris}}{1991a}]{h91araa}
{Harris}, W.~E. 1991a, ARA\&A, 29, 543

\bibitem[\protect\citeauthoryear{{Harris}}{{Harris}}{1991b}]{h91pasp}
{Harris}, W.~E. 1991b, PASP, 103, 32

\bibitem[\protect\citeauthoryear{{Harris}, {Harris}, \& {McLaughlin}}{{Harris}
  et~al.}{1998}]{HHM98}
{Harris}, W.~E., {Harris}, G. L.~S.,  \& {McLaughlin}, D.~E. 1998, AJ, 115,
  1801

\bibitem[\protect\citeauthoryear{{Harris} \& {Pudritz}}{{Harris} \&
  {Pudritz}}{1994}]{HP94}
{Harris}, W.~E.,  \& {Pudritz}, R.~E. 1994, ApJ, 429, 177

\bibitem[\protect\citeauthoryear{{Harris} \& {van~den~Bergh}}{{Harris} \&
  {van~den~Bergh}}{1981}]{HvdB81}
{Harris}, W.~E.,  \& {van~den~Bergh}, S. 1981, AJ, 86, 1627

\bibitem[\protect\citeauthoryear{{Kormendy} \& {McClure}}{{Kormendy} \&
  {McClure}}{1993}]{KM93}
{Kormendy}, J.,  \& {McClure}, R.~D. 1993, AJ, 105, 1793

\bibitem[\protect\citeauthoryear{{Lin} \& {Faber}}{{Lin} \&
  {Faber}}{1983}]{LF83}
{Lin}, D. N.~C.,  \& {Faber}, S.~M. 1983, ApJ, 266, L21

\bibitem[\protect\citeauthoryear{{McLaughlin} \& {Pudritz}}{{McLaughlin} \&
  {Pudritz}}{1996}]{MP96}
{McLaughlin}, D.~E.,  \& {Pudritz}, R.~E. 1996, ApJ, 457, 578

\bibitem[\protect\citeauthoryear{{Miller} et~al.}{{Miller}
  et~al.}{1998}]{de6data}
{Miller}, B.~W., {Ferguson}, H.~C., {Lotz}, J.~M., {Stiavelli}, M.,  \&
  {Whitmore}, B.~C. 1998, in preparation (Paper II)

\bibitem[\protect\citeauthoryear{{Miller} et~al.}{{Miller}
  et~al.}{1997}]{n7252}
{Miller}, B.~W., {Whitmore}, B.~C., {Schweizer}, F.,  \& {Fall}, S.~M. 1997,
  AJ, 114, 2381

\bibitem[\protect\citeauthoryear{{Moore}, {Lake}, \& {Katz}}{{Moore}
  et~al.}{1998}]{MLK98}
{Moore}, B., {Lake}, G.,  \& {Katz}, N. 1998, ApJ, 495, 139

\bibitem[\protect\citeauthoryear{{Sarajedini} \& {Layden}}{{Sarajedini} \&
  {Layden}}{1995}]{SL95}
{Sarajedini}, A.,  \& {Layden}, A.~C. 1995, AJ, 109, 1086

\bibitem[\protect\citeauthoryear{{Searle} \& {Zinn}}{{Searle} \&
  {Zinn}}{1978}]{SZ78}
{Searle}, L.,  \& {Zinn}, R. 1978, ApJ, 220, 357

\bibitem[\protect\citeauthoryear{{Tyson} \& {Scalo}}{{Tyson} \&
  {Scalo}}{1988}]{TS88}
{Tyson}, N.~D.,  \& {Scalo}, J.~M. 1988, ApJ, 329, 618

\bibitem[\protect\citeauthoryear{{Whitmore} et~al.}{{Whitmore}
  et~al.}{1995}]{WSLMB95}
{Whitmore}, B.~C., {Sparks}, W.~B., {Lucas}, R.~A., {Macchetto}, F.~D.,  \&
  {Biretta}, J.~A. 1995, ApJ, 454, L73

\end{thebibliography}

\clearpage

\begin{deluxetable}{llrrrrrr}
\tablenum{1}
\tablewidth{0pt}
\tablecaption{$S_N$ for Dwarf Elliptical Galaxies from {\it HST}}
\tablehead{
\ctlea{Galaxy} & \ctlea{Catalog} & \ctlea{$M_V^0$} & \ctlea{$N_c$\tablenotemark{(a)}} &  \ctlea{$S_N$} & \ctlea{$R_{\rm proj}$} & \ctlea{$\sigma_{\rm proj}$} &
\ctlea{$P$}\eol
\colhead{} & \colhead{Type}  & \colhead{} & \colhead{}&  \colhead{} & 
\colhead{[pc]} & \colhead{} & \colhead{[\%]}
}
\startdata
\sidehead{Nucleated Galaxies}
\cola VCC 1073\colc dE3N\cold $-17.4$\cole  $17.7\pm6.1$ \colg   $2.0
\pm   0.7$\coli   1.6 & 1.3 & $<10^{-5}$ \eol
\cola FCC \phn136\colc dE2N\cold $-17.2$\cole  $20.1\pm   5.6$\colg   $2.7
\pm   0.8$\coli   1.5 & 0.5 & 0.01\eol
\cola VCC 1876\colc dE5N\cold $-16.6$\cole  $24.7\pm   6.8$\colg   $5.5
\pm   1.6$\coli  14.3  & 1.5 & 0.2\phn \eol
\cola VCC 1254\colc dE0N\cold $-16.5$\cole  $24.6\pm   8.2$\colg   $6.2
\pm   2.2$\coli   2.2 & 1.2 & $<10^{-5}$\eol
\cola FCC \phn324\colc dS01(8)\cold $-16.4$\cole  $12.0\pm   4.8$\colg   $3.3
\pm   1.4$\coli   16.2 & 1.2 & 0.5\phn \eol
\cola FCC \phn150\colc dE4N\cold $-16.2$\cole   $7.4\pm   4.2$\colg   $2.4
\pm   1.4$\coli   4.8 & 2.3 & 0.01 \eol
\cola VCC \phn452\colc dE4N\cold $-15.8$\cole  $10.8\pm   5.8$\colg   $5.0
\pm   2.7$\coli   2.5 & 0.7 & 0.4\phn \eol
\cola FCC \phn174\colc dE1N\cold $-15.3$\cole   $9.8\pm   3.8$\colg   $7.2
\pm   2.9$\coli   11.8 & 1.7 & 0.2\phn\eol
\cola FCC \phn316\colc dE3N\cold $-15.2$\cole  $17.8\pm   5.7$\colg  $14.6
\pm   4.9$\coli   15.6 & 2.6 & 0.9\phn\eol
\cola FCC \phn254\colc dE0N\cold $-14.7$\cole   $5.7\pm   4.1$\colg   $7.7
\pm   5.7$\coli   25.0 & 0.8 & 0.01\eol
\cola VCC \phn503\colc dE3N\cold $-14.5$\cole   $1.7\pm   3.6$\colg   $2.8
\pm   6.0$\coli  9.2 & 0.7 & 0.2\phn\eol
\cola LGC \phn\phn50\colc dEN\cold $-14.0$\cole   $5.5\pm   3.1$\colg  $13.9
\pm   9.2$\coli  17.8  & 1.6 & 1.4\phn\eol
\cola VCC \phn240\colc dE2N\cold $-13.9$\cole   $8.6\pm   4.1$\colg  $23.6
\pm  14.0$\coli  30.1 & 0.8 & 0.4\phn\eol


\sidehead{Non-Nucleated Galaxies}

\cola VCC \phn\phn\phn9\colc dE1N\cold $-17.5$\cole  $22.9\pm   6.3$\colg   $2.3
\pm   0.7$\coli 155.0 & 11.0 & 6.0 \eol
\cola VCC \phn917\colc dE6\cold $-16.3$\cole   $6.8\pm   5.3$\colg   $2.1
\pm   1.6$\coli 267.9 & 160.0 & 20.0\eol
\cola VCC 1577\colc dE4\cold $-15.8$\cole  $14.9\pm   5.5$\colg   $7.3
\pm   2.8$\coli 34.4  & 1.6 & 3.0 \eol
\cola LGC \phn\phn47\colc dE\cold $-15.5$\cole   $3.5\pm   3.9$\colg   $2.2
\pm   2.4$\coli 572.8 & 64.0 & 70.0\eol
\cola VCC \phn118\colc dE3\cold $-15.3$\cole   $3.4\pm   3.7$\colg   $2.5
\pm   2.7$\coli 173.9 & 24.0 & 12.0\eol
\cola VCC 1762\colc dE6\cold $-15.2$\cole   $4.0\pm   3.7$\colg   $3.4
\pm   3.1$\coli  48.1 & 8.0 & 6.0 \eol
\cola FCC \phn110\colc dE4\cold $-14.9$\cole  $0.0\pm  3.8$\colg   $0.0
\pm   4.2$\coli 718.5 & 40.0 & 45.0\eol
\cola FCC \phn\phn48\colc dE3\cold $-14.8$\cole   $5.1\pm   4.4$\colg   $6.1
\pm   5.3$\coli  61.5 & 1.9 & 11.0\eol
\cola VCC 1651\colc dE5\cold $-14.6$\cole   $1.1\pm   3.9$\colg   $1.6
\pm   5.7$\coli 562.3 & 7.5 & 8.0\eol
\cola FCC \phn\phn64\colc dE5\cold $-14.2$\cole  $0.0\pm  2.6$\colg   $0.0
\pm   5.4$\coli 1257.0 & 60.0 & 50.0\eol
\cola VCC 2029\colc dE3\cold $-13.9$\cole   $1.1\pm   3.6$\colg   $3.1
\pm  10.4$\coli 124.6 & 5.1 & 7.0\eol
\enddata
\tablenotetext{(a)}{Corrected for undetected portion of the LF.}
\end{deluxetable}

\clearpage

\end{document}